\documentclass[11pt]{article}
\usepackage{amstex}
\usepackage{amssymb}
\input{psfig.tex}
\begin{document}	
\pagestyle{plain}
\hsize = 6.5 in 				
\vsize = 9.0 in		   
\hoffset = -0.75 in
\voffset = -0.95 in
\baselineskip = 0.22 in	

\centerline{\Large{\bf From Discrete Protein Kinetics to Continuous Brownian}}
\vskip 0.2 cm \noindent
\centerline{\Large{\bf Dynamics: A New Perspective}}
\vskip .5cm \noindent
\centerline{Hong Qian}  
\vskip 0.3cm \noindent
\centerline{Department of Applied Mathematics}
\centerline{University of Washington, Seattle, WA 98195-2420}
\vskip 0.5cm
\centerline{qian@amath.washington.edu}
\vskip 0.5cm
\centerline{\today}

\vskip 1.0cm \noindent

	Conformational fluctuation is a fundamental characteristic of 
proteins in aqueous solution, which differentiates the macromolecules
from small molecules.  The intrinsic beauty and the remarkable details
of the protein structures from crystallography often resulted in the 
view that proteins are static.  However, it is the conformational 
flexibility along with the well-defined structures which give 
rise to the versatile, almost magic, functionalities of proteins and
enzymes (Karplus and McCammon, 1983).  Ever since the conception of 
allosteric enzyme (Koshland et al., 1966, Monod et al., 1966), the 
multiple state notion of proteins has been widely appreciated.  Two 
particular notable examples are the models for {\bf folding kinetics}
of soluble globular proteins and the {\bf gating kinetics} 
of membrane channel proteins.  Both these models introduce discrete 
conformational states which are macroscopic and operationally defined 
by kinetic experiments (Tsong et al., 1972, Ehrenstein et. al., 1974).  
These models are usually expressed as
\begin{equation}
  U \overset{k_1}{\underset{k_2}{\rightleftharpoons}} N, 
	\hspace{1cm}
   O \overset{k_{\alpha}}{\underset{k_{\beta}}{\rightleftharpoons}} C
\label{2state}
\end{equation}
where $U$ and $N$ are unfolded and native states of a soluble
protein, with $k_1$ and $k_2$ as folding and unfolding rate constants.  
Similarly, $O$ and $C$ are open and closed states of a membrane channel 
protein, with $k_{\alpha}$ and $k_{\beta}$ as closing and openning rate 
constants.  When such simple models can not explain specific experiments,
usually more intermediate states are added (Bezanilla et al., 1994,
Baldwin, 1995).

	The discrete state description of proteins, however, neglects
conformational fluctuations within each state.  The energy landscape 
theory, treating the polypeptides as polymers, is an more realistic 
view of the protein dynamics (Frauenfelder et al., 1991, Wolynes et al., 
1995, Zwanzig, 1995, Doyle et al., 1997).  More importantly, 
recent experimental studies on several proteins indicate that 
it is also necessary to invoke continuous energy landscapes in 
order to provide comprehensive interpretations for the experiments
(Sigg et al., 1999, Qian and Chan, 1999).  There are 
now a host of laboratory observations which call attentions to
interpretations based on continuous energy.  Most notably are ($i$)
rapid early conformational changes in relaxation kinetics and ($ii$) 
nonactivated transitions induced by strong external forces.  
From a conceptual standpoint, as we shall show, these two types 
of observations are intimately related to the hysteresis and 
``bond rupturing'' phenomena recently observed in the receptor-ligand 
dissociation under atomic force microscopy (Florin et al., 1994; 
Moy et al., 1994; Evans and Ritchie, 1997, Shapiro and Qian, 1997, 1998, 
Merkel et al., 1999).  

	In the continuous approach, the kinetics of a molecule is 
viewed as a Brownian motion on an energy surface.  The theoretical basis 
of this approach has been extensively studied by Smoluchowski, 
Kramers, and others and is summarized in a review article
on the fifty years after Kramers theory (H\"{a}nggi et al., 1990).  
It is interesting to see that the merging of discrete with continuous 
kinetic models has been a long process in chemistry, and it is now 
the turn for proteins.  The following is a quotation from the
article, which gives us a historical perspective.
\begin{quote}
... For unimolecular gas phase reactions, a description of the rate 
in terms of \underline{discrete} energy exchange was more suitable 
than the continuous energy-exchange mechanism underlying energy
diffusion in Kramers' model (1941).  Work on chemical reactions
in condensed phase, for which the Kramers theory is most appropriate,
had to await the experimental progress achieved in the late seventies
and eighties. 
\end{quote}

	It is important to point out that every kinetic relaxation 
experiment has to involve \underline{two} conditions for the protein 
under study.  For protein folding they are usually the solvent condition 
or temperature, and for ion-channel gating it is membrane electrical
potential.  Recent experiments also apply external mechanical forces on 
proteins.  At time zero, the protein under condition 1 is subjected to 
condition 2, which initiates the relaxation kinetics of the molecule.  In 
terms of the energy function (potential of mean force), there are two 
\underline{different} energy functions corresponding to the two conditions 
(Karplus and Shakhnovich, 1992, Qian and Chan, 1999).
For the two-state kinetics show in Eq. \ref{2state}, the corresponding 
energy landscapes are shown in Fig. 1A.  Each discrete state is associated 
with an energy well.  The continuous model, however, also points out that 
the shape of an energy well, as well as their relative heights, change 
with the condition.  Immediately after the initiation, 
not only the thermal equilibrium between the two wells are perturbed,
the equilibrium within each of the wells too is perturbed.   Therefore 
the response of a protein to the perturbation is to readjust its equilibrium 
distributions within each energy wells, as well as to redistribute 
between the two wells.  Since the latter process is thermally activated 
while the former process is energetically down-hill, the readjustment 
contributes a rapid early conformational change in any relaxation kinetics 
of proteins.  One unique feature of this process, however, is that it is 
not thermally activated, hence non-exponential.  Experimentally, from 
such non-exponential diffusion process one is expected to observe faster 
relaxation time with faster measuremental temporal resolution, reminiscent 
of a fractal behavior.

	Since the down-hill readjustment within each energy well is 
usually much faster than the two-state barrier crossing, and since 
the magnitude of the readjustament is usually small relative to that
of the two-state transition, such early kinetic events are difficult 
to observe experimentally.  However, recent experiments on gating 
of voltage-dependent membrane ion channel proteins have observed such
fast kinetic phase (Stefani and Bezanilla, 1996), and a Brownian 
diffusion mechanism has been proposed for the early fast ($\sim \mu$s) 
components in the movement of gating-charge of the channel responding 
to a sudden change in membrane voltage (Sigg et al., 1999). 
In the kinetics of protein folding, such a fast energetically down-hill 
event is also observed (Sosnick et al., 1996, Hagen et al., 1996,
Qi et al., 1998).  In addition, a large amount of experimental observation 
of various intermediate states in the early time of protein folding 
kinetics (known as molten globular states) can be interpreted by a 
readjustment step responding to a sudden change in denaturant 
concentration in solvent (Karplus and Shakhnovich, 1992,
Qian and Chan, 1999).  

	A second situation under which discrete kinetics fail
to provide a cogent interpretation is when the perturbation is so 
large that it completely eliminates the activation barrier, as shown 
in Fig 1B.  Under such conditions, the traditional rate process with
thermal activation loses its meaning all together and
the relaxation is a very fast energetically down-hill diffusion.
This phenomenon has not been observed in monomeric protein folding 
(unfolding) kinetics.  However, such a mechanism lies behind 
streptavidin-biotin bond rupturing with atomic force microscopy 
(Shapiro and Qian, 1997, 1998), as well as successive unfolding of 
giant muscle protein titin by force (Qian and Shapiro, 1999).  
A similar behaviour also has been observed 
in the gating kinetics of $K^+$ channel with an extreme holding 
potential (Sigg et al., 1999). 

	The introduction of continuous energy landscape does not
invalidate the discrete transition between two energy wells, rather it
generalizes the discrete model with an increasing molecular details.  
It is well known that with a sufficient
large activation barrier separating two energy wells, there is a rapid 
equilibrium within each well. Furthermore, the transition from one well 
to another is essentially exponential, i.e., Arrhenius (also known as
discrete-state Markovian).  This is the theoretical basis for the practice 
of discrete-state kinetics.  One should recognize, however, that the
``molecular structures'' of the discrete states, which usually are 
defined experimentally through spectroscopy, change with the environmental
condition for the protein.  These changes also reflected in the 
baselines when fitting discrete multi-state models to equilibrium 
measurements (Qian, 1997, Qian and Chan, 1999). 

	Another important reason for introducing the continuous energy 
function to augement the discrete-state kinetics is the inability of 
relating energy to force in the latter framework.  Force is the change 
of energy in response to a change in distance.  As we can see the 
concept of distance is completely missing in the discrete-kinetics.  
In the continuous energy landscape, no matter how ill-defined the 
reaction coordinates are, they provides a conceptual framework.  
Therefore, the continuous energy landscape provides a bridge 
between the experimental studies on kinetic of proteins 
and more direct measurements of force and displacement on
single protein molecules (Kellermayer et al., 1997, Reif et al., 1997,
Tskhovrebova et al., 1997, Qian and Shapiro, 1999).

	In summary, one important consequence of the energy landscape 
concept is that within each discrete kinetic state, there could be 
significant conformational readjustment due a changing condition 
(perturbation) for the protein, such as changes in temperature, 
solvent, or membrane potential.  Therefore, following a sudden change 
in one of these conditions, a protein has two characteristic kinetic 
steps: an energetically down-hill readjustment into the new equilibrium 
position within the same discrete state, and then an thermally 
activated rate process which jumps from one discrete state into another 
with lower energy.  When the perturbation is sufficiently larger, it is 
also possible that the activation barrier is completely eliminated.
Then the kinetics becomes a down-hill diffusion, 
and relaxation kinetics is no longer exponential.  The continuous 
energy perspective on protein kinetics provides a comprehensive 
theoretical framework for a host of experimental observations, ranging
from protein folding, to membrane channel gating, to protein-ligand
dissociation and protein unfolding under external force.

	The conceptual thrust of the continuous energy landscape approach 
to proteins is that it provides a theoretical language for discussing a
wide range of dynamical behavior of proteins.  It has laid a foundation
for developing a macromolecular mechanics at a mesoscopic level
between the discrete models and the atomic-level molecular dynamics 
(Qian and Shapiro, 1999).  It allows important concepts such as force and 
movement to be discussed on an equal footing as energy and thermodynamic
states.  With the recent significant progress in biophysical measurements   
of forces and movements in single protein molecules, models based on
continuous Brownian dynamics will become an indispensable 
part of the protein science.  In a similar spirit, Eisenberg
and his colleagues have developed a diffusion theory for ion
movement (not to be confused with protein movement in the gating
kinetics) in open channels to augement traditional discrete-state 
models.  For a review see Cooper et al. (1988) and Eisenberg (1996).

	 Here is an example to show how the continuous energy 
function serves as a unifying theoretical edifice in molecular
biophysics.  One interesting phenomenon observed from protein-ligand
interaction under external force is hysteresis: the 
association process under a force and the dissociation
process under a force are significantly different.  This
can be quantitatively interpreted as in Fig. 2.  Compare
this model with the well-known protein folding-unfolding
kinetics scheme below, one immediately sees that the 
main feature of the two molecular processes are indeed
identical.   

\vskip 0.5cm
\[ \begin{array}{c|cccc}
    & &\textrm{unfolded state} & &\textrm{folded state}  \\
    \hline & & & & \\
    \textrm{native condition} &\hspace{0.5cm} &\textrm{wet molten globule} 
    	&\Longrightarrow  &\textrm{native structure}  \\
    & & & &  \\ & &\uparrow & &\downarrow \\ & & & & \\
    \textrm{denaturing condition} & &\textrm{random coil} 
	& \Longleftarrow &\textrm{dry molten globule} \\
    & & & & \\ \hline
     \end{array}   \] 
\vskip 0.5cm \noindent
where we introduce the term ``wet molten globule'' referring the collapsed
intermediate state commonly observed in the protein refolding kinetics
(cf. Baldwin, 1993).  It has been interpreted as the unfolded state 
under a native condition (Dill and Shortle, 1991, Qian and Chan, 1999).  
The dry molten globule (Kiefhaber et al., 1995), on the other hand, has 
been interpreted as the folded state under a denaturing condition 
(Qian and Chan, 1999).  Both these two states are kinetic intermediates 
which appear in the transient folding and unfolding processes, respectively. 
The wet molten globule is a kinetically metastable state before the 
major activation barrier in the \underline{folding} process, while 
the dry molten globule is again a kinetically metastable state 
before the major activation barrier but in the \underline{unfolding}
process.  This difference is the same as the hyseresis.

\baselineskip = 0.17 in	
\vskip 1.0cm \noindent
{\bf References}
\vskip 0.5cm \noindent
\def \SN{\item{}}
\small
\begin{description}
\SN Bezanilla, F., Perozo, E., and Stefani, E. (1994) {\it Biophys. J.}
66, 1011-1021.  
\SN Baldwin, R.L. (1993) {\it Curr. Opin. Struct. Biol.} 3, 84-91.  
\SN Baldwin, R.L. (1995) {\it J. Biomol. NMR} 5, 103-109. 
\SN Cooper, K.E., Gates, P.Y., and Eisenberg, R.S. (1988)
{\it J. Membrane. Biol.} 106, 95-105. 
\SN Dill, K.A. and Shortle, D. (1991) {\it Ann. Rev. Biochem.} 60, 795-825.
\SN Doyle, R., Simons, K., Qian, H., and Baker, D. (1997) {\it Proteins:
Struct. Funct. Genet.} 29, 282-291.
\SN Ehrenstein, G., Blumenthal, R., Latorre, R., and Lecar, H.
(1974) {\it J. Gen. Physiol.} 63, 707-721.   
\SN Eisenberg, R.S. (1996) {\it J. Membrane Biol.} 150, 1-25.     
\SN Evans, E., and Ritchie, K. (1997) {\it Biophys. J.} 72, 1544-1555.
\SN Florin, E., Moy, V.T., and Gaub, H.E. (1994) {\it Science}, 264, 415-417. 
\SN Frauenfelder, H., Sligar, S.G., and Wolynes, P.G. (1991) {\it Science}
254, 1598-1603.  
\SN Hagen, S.J., Hofrichter, J., Szabo, A., Eaton, W.A. (1996) {\it Proc.
Natl. Acad. Sci. USA} 93, 11615-11617. 
\SN H\"{a}nggi, P., Talkner, P., and Borkovec, M. (1990) {\it Rev.
Mod. Phys.} 62, 251-341.  
\SN Kramers, H.A. (1941) {\it Physica} 7, 284-304.
\SN Karplus, M. and McCammon, J.A. (1983) {\it Ann. Rev. Biochem.}
52, 263-300.
\SN Karplus, M. and Shakhnovich, E. (1992) In {\it Protein Folding.} 
Creighton, T.E. Ed., W.H. Freeman, New York.  
\SN Kellermayer, M.S.Z., Smith, S.B., Granzier, H.L., and
Bustamante, C. (1997) {\it Science} 276, 1112-1116.
\SN Kiefhaber, T., Labhardt, A.M., and Baldwin, R.L. (1995) 
{\it Nature} {\bf 375}, 513-515.
\SN Koshland, D.E., Nemethy, G., and Filmer, G. (1966) {\it Biochem.} 
5, 365-385. 
\SN Merkel, R., Nassoy, P.,  Leung, A., Ritchie, K., and Evans, E.
(1999) {\it Nature}, 397, 50-53.
\SN Monod, J., Wyman, J., and Changeux, J.-P. (1965) {\it J. Mol. Biol.}
12, 88-118. 
\SN Moy, V.T., Florin, E., and Gaub, H.E. (1994) {\it Science}, 266, 257-259. 
\SN Qi, P.X., Sosnick, T.R., and Englander, S.W. (1998) {\it Nature Struct.
Biol.} 5, 882-884.  
\SN Qian, H. (1997) {\it J. Mol. Biol.} 267, 198-206.
\SN Qian, H. and Chan, S.I. (1999) {\it J. Mol. Biol.} 286, 607-616. 
\SN Sigg, D., Qian, H., and Bezanilla, F. (1999) {\it Biophys. J.}
76, 782-803.  
\SN Reif, M., Gautel, H., Oesterhelt, F., Fernandez, J.M., and
Gaub, H.E. (1997) {\it Science} 276, 1109-1112.
\SN Shapiro, B.E. and Qian, H. (1997) {\it Biophys. Chem.}
67, 211-219.
\SN Shapiro, B.E. and Qian, H. (1998) {\it J. Theoret. Biol.}
194, 551-559.
\SN Sosnick, T.R., Mayne, L., and Englander, S.W. (1996) {\it Proteins:
Struct. Funct. Genet.} 24, 413-426.  
\SN Stefani, E. and Bezanilla, F. (1996) {\it Biophys. J.}
70, A134 (Abstr). 
\SN Szabo, A., Schulten, K., and Schulten, Z. (1980) {\it J. Chem. Phys.}
72, 4350-4357.  
\SN Tsong, T.Y., Baldwin, R.L., McPhie, P., and Elson, E.L. (1972) {\it J.
Mol Biol.} 63, 453-475. 
\SN Tskhovrebova, L., Trinick, J., Sleep, J.A., and Simmons, R.M. (1997)
{\it Nature} 387, 308-312.
\SN Wolynes, P.G., Onuchic, J.N., and Thirumalai, D. (1995) {\it Science},
267, 1619-1620. 
\SN Zwanzig, R. (1995) {\it Proc. Natl. Acad. Sci. USA} 92, 9801-9804.
\end{description}

\normalsize
\baselineskip = 0.22 in	

\vskip 1.0cm \noindent
{\bf Figure Captions} 

\vskip 0.3cm \noindent
\underline{Figure 1} Schematic diagram showing continuous energy 
functions for a two-state protein kinetics (Eq. \ref{2state}) under 
two different conditions. (A) It is shown that the shape of each 
energy well, as well as the relative heights of the two wells, 
change with the condition.  (B) Under extreme condition, the 
activation-energy barrier can completle disappear. In this case,
the relaxation process is an energetically down-hill diffusion.

\vskip 0.3cm \noindent
\underline{Figure 2}  When applying a force to a protein-ligand
pair, one can either pull the ligand apart from the protein, or 
load the protein with the ligand.  The events in these two
kinetics processes are schematically shown here.  The particle
in the diagram represents the ligand which experiences both 
intermolecular force from the protein (modeled as a 6-12 potential)
and the force probe (modeled as a Hookean spring).  The energy 
minimum on the left is the equilibiurm position for the intermolecular 
energy; while the energy minimum on the right is the equilibrium 
position of the spring.  For more discussion, see Qian and Shapiro, 
1999.

\vskip 2cm \noindent
Running Title: Discrete Kinetics and Continuous Dynamics 


\pagebreak

\begin{figure}[h]
\[
\psfig{figure=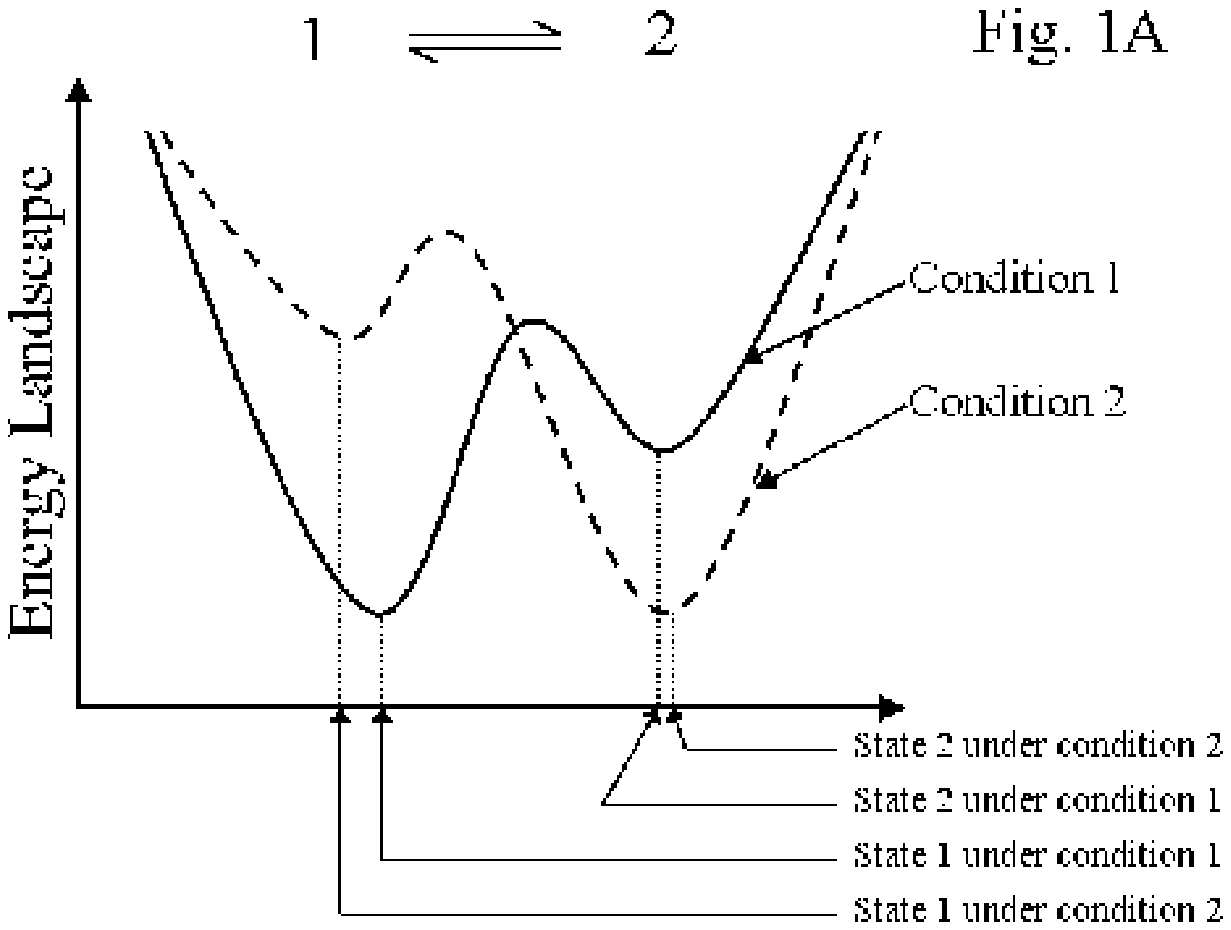,%
width=5.5in,height=4.5in,%
bbllx=1.5in,bblly=3.5in,%
bburx=7.in,bbury=8in}
\]
\vskip 1.5cm
\[
\psfig{figure=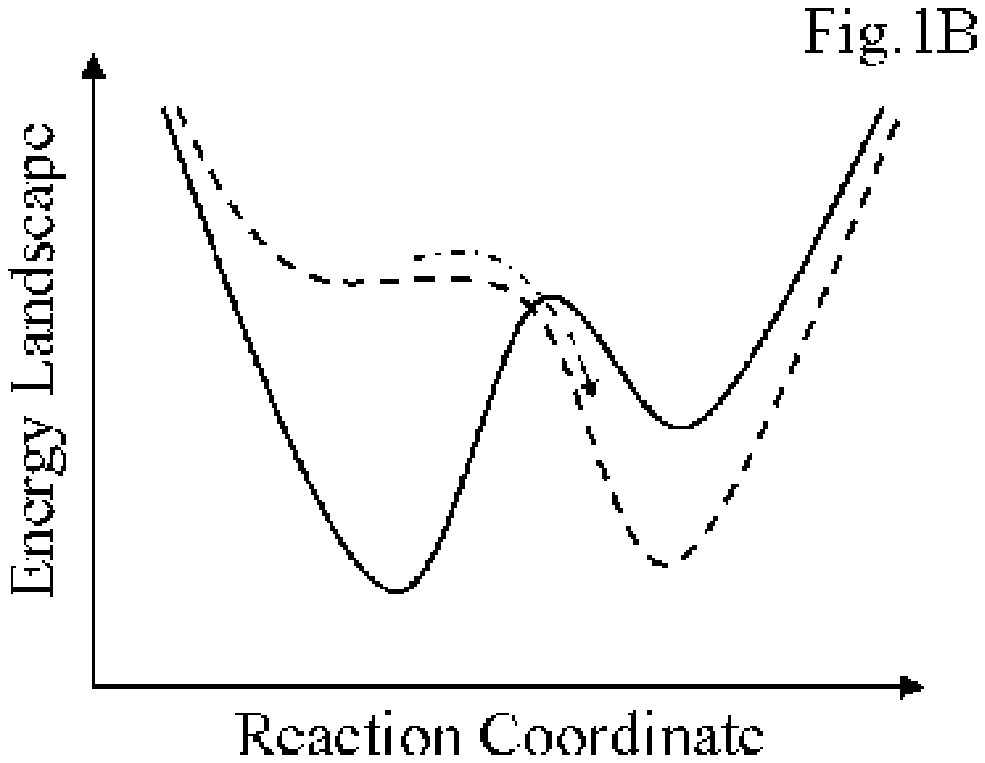,%
width=6.0in,height=3in,%
bbllx=1.5in,bblly=4in,%
bburx=7.5in,bbury=7in}
\]
\end{figure}

\begin{figure}[h]
\[
\psfig{figure=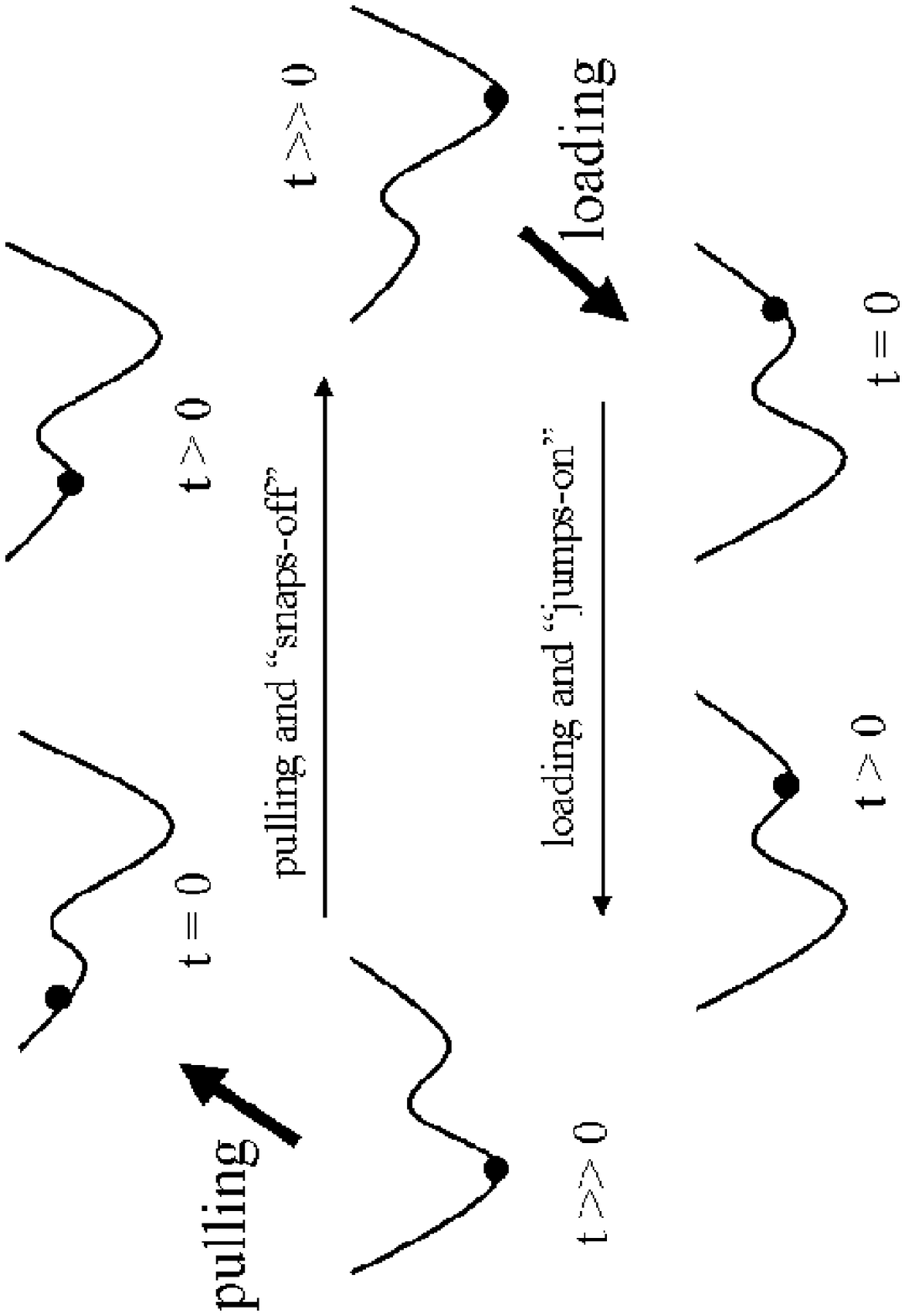,%
width=5.5in,height=7.5in,%
bbllx=0.5in,bblly=0.5in,%
bburx=8.in,bbury=10.5in}
\]
\end{figure}

\end{document}